# Ethics Suppression in AI Development: A Comparative Case Study of Responsible AI Under Competitive Pressure

**Authors:**

Victor Frimpong
SBS Swiss Business School, Kloten-Zurich, Switzerland
ORCiD: 0009-0004-9560-3927

**Abstract:**

This study explores the governance tensions among Google, OpenAI, Meta, and Microsoft, focusing on how competitive AI development can lead to the ethics suppression. The study aims to examine how ethical oversight loses practical influence even when responsible AI structures are formally in place, especially during rapid technological competition. Despite the rapid growth of responsible AI frameworks and ethical principles in the tech sector, there has been less focus on how these concerns are actually applied when pressure to deploy increases. This paper introduces the concept of ethics suppression in AI development, highlighting that while ethical oversight may exist on paper, its practical authority can diminish during decision-making. Using a comparative qualitative case study examining corporate governance documents, public statements, and reports on governance issues, we identify common issues such as reduced authority, a weakened ability to escalate concerns, urgency in deployment, and fragmented governance. Our findings indicate that heightened competition can weaken the influence of ethical oversight, even in organisations with established responsible AI practices. By differentiating ethics suppression from ethics washing, it offers a framework for organisations and policymakers to evaluate the effectiveness of ethical oversight in competitive AI development.

## Introduction

AI governance has seen rapid growth recently, with governments, international organisations, technology companies, and academic institutions creating frameworks that emphasise responsible AI, transparency, fairness, accountability, and safety. The introduction of regulatory initiatives such as the European Commission's AI Act highlights the growing adoption of ethical AI governance worldwide (Papagiannidis et al., 2025).

AI development environments are now highly competitive due to the fast commercialisation of generative AI, intensified competition among tech companies, and the demand for faster deployment cycles. Organisations must innovate quickly while ensuring ethical oversight, safety, and governance accountability. (Papagiannidis et al., 2025).

The ongoing tension highlights a significant but underexplored issue in responsible AI research. While extensive studies have covered AI ethics principles, governance frameworks, fairness, explainability, and regulatory models (Corrêa et al., 2023; Papagiannidis et al., 2025), less attention has been paid to how ethical concerns lose their impact over the course of AI development. Recent controversies involving internal dissent, governance changes, reduced safety oversight, and disagreements over deployment priorities illustrate that having ethical principles does not necessarily ensure their influence on organisational decision-making (Harvard Law Review, 2025; Tan, 2024).

This paper introduces "ethics suppression" in AI development, which is not about intentional censorship but rather the weakening or marginalisation of ethical concerns under competitive pressure. In these conditions, ethical issues, safety objections, and governance processes may be sidelined, even if organisations maintain a public commitment to ethical principles and frameworks. This results in ethics losing its influence in the actual development processes.

The paper argues that competitive pressure can create tensions between responsible AI governance and deployment incentives. As development cycles shorten, ethical review processes may lose authority and be bypassed in favour of speed and market competitiveness. This isn't necessarily a rejection of ethics by organisations, but rather the result of competitive dynamics that undermine ethical oversight during AI development.

To investigate this issue, the paper employs a comparative qualitative case study approach, examining major AI organisations that claim to promote responsible governance yet face intense competition. This study aims to explain how ethics suppression occurs in competitive AI development environments and create a framework to analyse the impact of competitive pressure on ethical oversight. By exploring governance tensions within major AI organisations, the research seeks to clarify the conditions under which formal ethical governance loses its operational influence, even if it remains officially visible.

## Literature Review

### The Expansion of AI Ethics and Responsible AI Governance

The swift adoption of artificial intelligence has prompted considerable concerns surrounding fairness, accountability, transparency, privacy, bias, safety, and the need for human oversight. In light of these challenges, various stakeholders, including governments, international organisations, corporations, and academic institutions, have developed comprehensive ethical frameworks to guide the responsible development and deployment of AI systems. Noteworthy initiatives in this area include the OECD AI Principles, the NIST AI Risk Management Framework, UNESCO's Recommendation on the Ethics of Artificial Intelligence, and the European Union AI Act. These frameworks emphasise key principles such as transparency, accountability, explainability, robustness, and human-centred governance (Jobin et al., 2019; OECD, 2019; NIST, 2023).

This expansion has facilitated the institutionalisation of responsible AI governance across both public and private sectors. Leading technology firms have established AI ethics boards, responsible AI teams, governance committees, internal review processes, and safety frameworks to operationalise ethical oversight within AI development environments (Morley et al., 2020). Consequently, the responsible AI literature has increasingly focused on designing governance mechanisms that effectively manage AI-related risks while fostering innovation and maintaining public trust.

Scholarly research has highlighted significant fragmentation in the governance of AI ethics. Ethical principles are often broad, abstract, and applied inconsistently across organisations and jurisdictions (Mittelstadt, 2019). While many frameworks align around common ethical values, translating these principles into effective operational practices presents substantial challenges. There has been a notable shift in focus from merely identifying ethical principles to a deeper examination of how governance structures operate within organisational contexts.

### From Ethical Principles to Operational Governance

Many ethical frameworks lack enforcement, accountability, and integration into development processes (Metcalf et al., 2019). This has sparked interest in responsible AI governance, with a focus on embedding ethical oversight within organisational systems and workflows. Research on responsible AI governance has identified several key mechanisms, including algorithmic auditing, model documentation, human oversight, impact assessments, ethics reviews, and risk classification systems (Raji et al., 2020). Scholars are also investigating interdisciplinary governance structures that combine legal, technical, ethical, and organisational viewpoints in AI development.

Translating ethical governance into effective operational influence is a significant challenge. Research shows that ethical review mechanisms often face ambiguity, limited authority, unclear escalation pathways, and competing commercial incentives (Greene et al., 2019). As a result, ethical governance may be formally established but is frequently not well integrated into product development decisions. This gap between the formal existence of ethical structures and their practical influence is a critical issue in the responsible AI discourse.

## Ethics Washing and Symbolic Governance

The literature on ethics washing offers a critical framework for examining the discrepancies between publicly stated ethical commitments and actual organisational practices. Ethics washing refers to instances where organisations adopt ethical language, principles, or governance structures primarily to project an image of responsibility, legitimacy, or regulatory compliance, without making substantial changes to their operational behaviours (Bietti, 2020).

Several scholars contend that voluntary AI ethics frameworks may serve a symbolic purpose, as they can alleviate external pressures while allowing organisations to sidestep more stringent regulatory measures or accountability methods (Hao, 2020). As a result, these ethics initiatives may bolster organisational legitimacy without necessarily imposing limitations on commercial or technological ambitions. This issue is particularly salient in the technology sector, where companies often publish comprehensive responsible AI principles even as they face scrutiny regarding their deployment practices, surveillance issues, data governance, labour practices, and safety risks.

The ethics washing literature highlights the gap between ethical representation and organisational implementation. However, it mainly focuses on external signalling and reputation management, overlooking the internal dynamics that can weaken ethical considerations during AI development. As a result, current research offers limited insights into how ethical oversight functions under rapid innovation and competitive pressures.

## Competitive Pressure and the Acceleration of AI Development

Recent advancements in generative AI have heightened competition among major technology firms, startups, and state-supported initiatives. The swift emergence of large language models and foundational technologies has incentivised organisations to expedite deployment, enhance market positioning, and achieve technological leadership and first-mover advantage (Bommasani et al., 2021). This landscape is characterised by rapid progress, confidentiality, high investment expectations, and strategic rivalry.

Research in innovation indicates that competitive pressure can significantly influence organisational decision-making, risk tolerance, and governance practices (Christensen, 1997). In environments defined by accelerated competition, organisations may prioritise timely product releases and market capture, potentially at the expense of more deliberate processes. Consequently, this dynamic can create tension between the pursuit of responsible governance and the imperatives of commercial deployment in AI.

Recent controversies highlight growing tensions within AI organisations. Cases involving the departures of prominent AI ethics researchers at Google, governance disputes and safety-related resignations at OpenAI, concerns about governance controls around Meta's Llama releases, and debates over the pace of generative AI deployment at Microsoft illustrate the difficulties of maintaining effective ethical oversight amid accelerated technological competition. While research recognises these tensions, there is a lack of empirical examination into how ethical concerns diminish in operational influence.

## Organisational Ethics and Ethics Suppression

Research in organisational ethics highlights how ethical issues can be overlooked in high-pressure environments. Concepts like organisational silence, ethical fading, and moral disengagement show that performance pressures, hierarchy, or institutional incentives can overshadow ethical concerns (Bird & Waters, 1989; Tenbrunsel & Messick, 2004).

Ethical fading occurs when moral considerations recede from attention because organisational actors increasingly focus on competing objectives such as efficiency, innovation, growth, or market competition (Tenbrunsel & Messick, 2004). Organisational silence indicates that employees may avoid raising ethical concerns if they believe escalation mechanisms are ineffective or risky. Ethical issues may lose influence not only through explicit suppression but also through institutional structures, incentive systems, authority distributions, or operational priorities.

Organisational ethics concepts are important but underdeveloped in AI governance research. Most existing AI ethics studies focus on principles, technical fairness, explainability, and governance structures, neglecting the organisational factors that influence ethics during development.

## Why Ethics Suppression Matters

Research on AI ethics and governance often overlooks the consequences of weakened ethical authority within organisations. When ethical oversight loses influence, the impact goes beyond governance, affecting organisational performance, legal risks, innovation, stakeholder trust, and long-term legitimacy. This highlights that ethics suppression is a critical governance issue, not just an internal management problem (Metcalf et al., 2019; Greene et al., 2019; Bietti, 2020).

Weakening ethical oversight can hinder internal accountability, create decision-making opacity, and decrease an organisation's ability to manage AI-related risks proactively. As governments implement more stringent AI regulations, ineffective governance may heighten regulatory exposure, leading to compliance failures and financial penalties (NIST, 2023; Papagiannidis et al., 2025).

While reducing governance friction may speed up product development and enhance short-term competitiveness, these benefits can be outweighed by increased compliance costs, deployment delays, and damage-control efforts following governance failures. Weaker oversight can allow biases and safety risks to persist, raising technical and operational vulnerabilities (Raji et al., 2020; Morley et al., 2020).

Diminished ethical governance can stifle innovation and erode public trust. Sustainable AI innovation relies not only on technological advancement but also on stakeholder confidence in responsible development. Governance failures or apathy towards ethics can harm an organisation's legitimacy, reduce consumer and investor trust, and weaken societal confidence in AI technologies (Jobin et al., 2019; OECD, 2019; UNESCO, 2021). Therefore, the effects of ethics suppression have far-reaching implications for accountability, responsible innovation, and the overall legitimacy of AI governance.

### Research Gaps

Current literature covers AI ethics principles, governance frameworks, and the issue of ethics washing. However, it lacks insight into how ethical concerns and governance can be undermined amid rapid innovation and competitive pressure. This paper introduces the concept of ethics suppression in AI development. Instead of just assessing whether organisations have ethical frameworks, it focuses on the conditions that affect the effectiveness of ethical oversight in AI development and deployment.

## Conceptual Framework

### Ethics Suppression in AI Development

This paper defines ethics suppression in AI development as a scenario where ethical oversight is formally in place but has little practical impact on decision-making. This situation occurs when authority, timing, escalation capacity, or organisational priorities shift against the ethical review process. Ethics suppression does not mean the absence of ethical governance or active hostility toward ethics; rather, it indicates that, while ethical review mechanisms exist, they are increasingly ineffective at influencing operational outcomes. This definition aligns with research showing that ethical concerns can diminish in importance when faced with competing priorities and institutional constraints (Bird & Waters, 1989; Tenbrunsel & Messick, 2004), as well as governance studies that differentiate between formal structures and their actual effectiveness (Meyer & Rowan, 1977; Metcalf et al., 2019).

This definition clarifies how the ethics suppression differs from similar concepts. Governance failure occurs when governance mechanisms are ineffective or poorly designed, failing to meet their goals. Governance weakening broadly refers to any decline in the effectiveness or authority of governance. In contrast, ethics suppression is a specific situation where ethical oversight is formally established but loses influence during development processes. This concept highlights the disparity between the existence of institutional oversight and its practical authority, building on previous discussions of organisational decoupling and symbolic governance in AI development (Meyer & Rowan, 1977; Bietti, 2020).

## Competitive Pressure and Ethical Influence

Competitive pressure plays a key role in shaping ethical practices during AI development. Organisations in fast-paced AI markets face strong incentives to accelerate deployment, secure investment, maintain leadership, and gain first-mover advantages. As a result, they may prioritise rapid growth and market positioning over thorough governance processes.

This creates tensions between responsible AI governance and operational priorities. Ethical reviews, safety evaluations, and escalation mechanisms can become viewed as hindrances to strategic goals rather than essential parts of the development process. Consequently, while ethics are not outright rejected, their influence may diminish due to organisational routines, tight timelines, changing authority structures, and a focus on commercial objectives.

The framework views the ethics suppression as a structural and organisational issue, not just an ideological one. Ethical concerns may diminish in impact not because organisations actively reject ethics, but because development environments prioritise speed, deployment, and competitiveness.

Figure 1 illustrates the proposed framework for ethics suppression in AI development. It shows how rising competitive pressure can diminish the effectiveness of ethical oversight, even when responsible AI structures are in place. The framework highlights the key organisational mechanisms that can lead to ethics suppression amid rapid AI advancement.

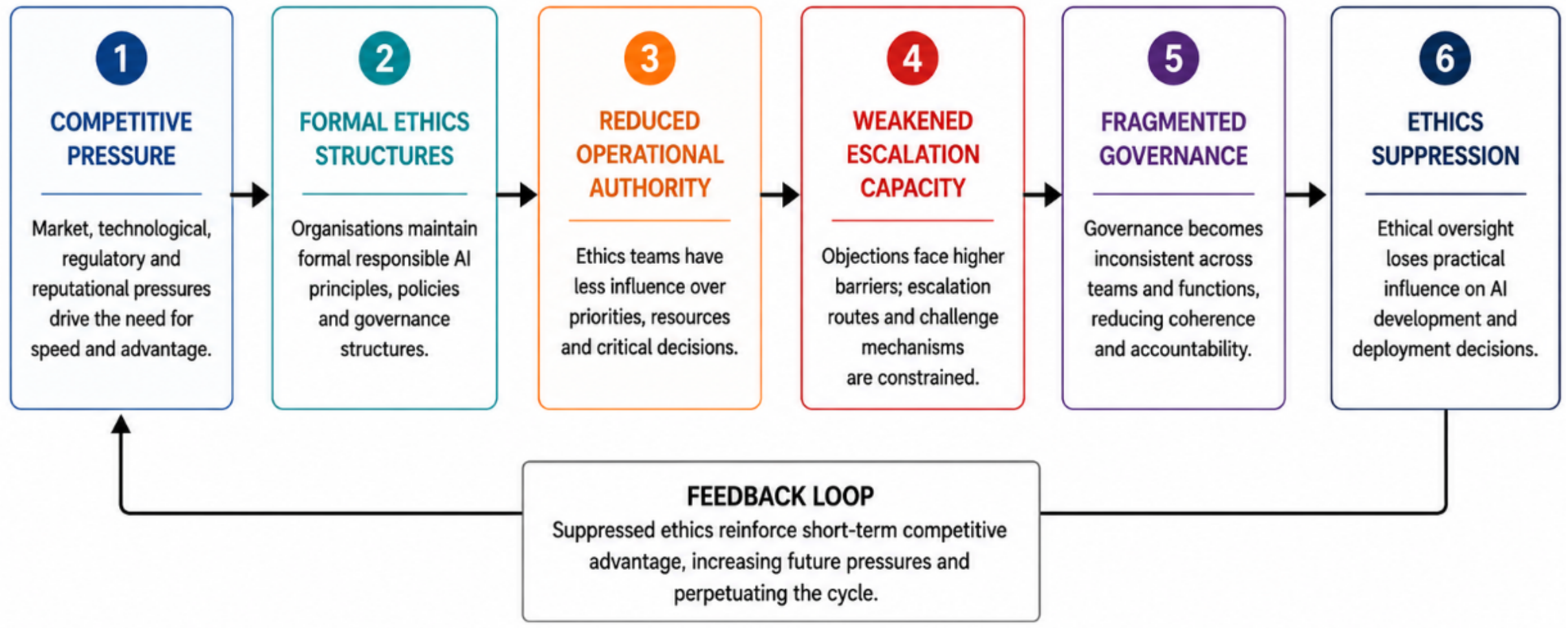

Figure 1: Conceptual framework of ethics suppression in competitive AI development
Source: The Author, 2026

The framework proposes that organisations may continue to maintain visible responsible AI policies, ethics principles, and governance structures while simultaneously experiencing a decline in their operational influence. As competitive pressure intensifies, ethical oversight may be progressively constrained by reduced operational authority, weakened escalation capacity, and fragmented governance, ultimately resulting in ethics suppression. The framework serves as an analytical model for examining how organisational conditions shape the effectiveness of ethical governance across the comparative case studies.

## Operational Mechanisms of Ethics Suppression

The framework identifies various organisational mechanisms that can lead to the suppression of ethical practices during AI development.

**(a) Reduced Escalation Capacity**

Ethical concerns can become irrelevant if governance structures lack the power to challenge or change development decisions. As a result, ethics teams may only serve in an advisory role, limiting their impact on deployment outcomes.

**(b) Temporal Compression**

Faster development cycles can limit time for ethical discussions, safety checks, and interdisciplinary reviews. This can lead to governance processes that focus more on procedures than on meaningful substance.

**(c) Strategic Prioritisation:**

Organisations often prioritise deployment speed and technological advancement due to commercial incentives and competitive pressures, which can push ethical concerns aside.

**(d) Organisational Fragmentation:**

AI governance typically involves various overlapping roles—engineering, legal, policy, compliance, security, product management, and leadership. This fragmentation can create unclear accountability and weaken coordinated ethical oversight.

**(e) Institutional Visibility Without Operational Authority:**

Organisations may project strong ethical commitments through public statements and reports, but their internal influence on decision-making can be limited. As a result, ethical principles may have only symbolic value rather than real authority.

These factors illustrate how ethical governance can weaken in organisations that publicly advocate for responsible AI.

### Ethics Suppression as an Organisational Process

The framework views the concept of ethics suppression as a dynamic process rather than a fixed state. Ethical influence can fluctuate based on changing strategic conditions, leadership priorities, governance authority, public scrutiny, regulatory pressure, and competitive intensity.

This approach differs from ethics washing literature, which focuses on the gap between external ethical signals and actual practices. Instead, this framework examines the internal processes that diminish the influence of ethical concerns during development, shifting the focus from public representation to decision dynamics within organisations.

It challenges traditional compliance-based governance models, which assume that formal structures ensure control. It highlights that governance authority can vary significantly depending on institutional conditions and competitive pressures.

### Analytical Model

The conceptual framework suggests that increasing competitive pressure and incentives for rapid deployment may weaken the role of ethical oversight in AI development. This does not mean that outcomes are predetermined; some organisations can maintain strong ethical authority despite such pressures, while others may see a decline in governance. The framework emphasises the importance of analysing how organisational conditions impact the effectiveness of responsible AI governance. It focuses on whether ethical structures can influence decisions in high-pressure environments, rather than merely on their existence. This distinction serves as the basis for the subsequent comparative case study analysis.

## Methods and Data

### Research Design

This study employs a comparative qualitative case study approach to examine how competitive pressures can overshadow ethical concerns in AI development. A qualitative case study design is suitable for exploring organisational processes, governance dynamics, institutional tensions, and decision-making structures in real-world settings, rather than focusing on specific causal relationships (Yin, 2018).

Comparative case study methods effectively analyse complex organisational phenomena influenced by various institutional, strategic, and governance factors (Stake, 1995; Eisenhardt, 1989). This approach enables a contextual exploration of responsible AI governance across different organisations, highlighting common patterns and mechanisms that suppress ethics.

Consistent with the overall research aim, this comparative qualitative case study investigates the emergence of ethics suppression as a governance issue in competitive AI development. With limited research on the decline of ethical authority in responsible AI governance, this design enables an in-depth examination of organisational processes across multiple cases, helping to develop and refine the proposed conceptual framework.

The unit of analysis in this study is the organisational governance processes through which ethics suppression emerges within AI development environments. Instead of looking at individual employees, specific AI systems, or isolated incidents, we examine the governance structures, decision-making processes, and institutional responses that influence ethical oversight under competitive pressure. The analysis uses four organisations as comparative cases to explore these governance dynamics.

## Case Selection

The study uses purposive case selection, targeting major AI organisations under significant competitive pressure that also uphold responsible AI governance. Cases were chosen based on four criteria:

1. significant involvement in advanced AI development.

2. publicly articulated responsible AI or AI ethics frameworks.

3. documented governance controversies, internal dissent, or ethical disputes related to AI deployment.

4. substantial exposure to competitive pressures associated with accelerated AI commercialisation.

Based on these criteria, the study concentrates on the following organisations:

• Google

• OpenAI

• Meta

• Microsoft

These organisations are key players in AI development, focusing on responsible governance while facing intense market competition and pressure to deploy rapidly. They also offer ample public documentation on ethics controversies, governance changes, safety issues, and internal conflicts.

Table 1 provides a clear summary of the main evidence sources, governance tensions, and analytical relevance for each case. It shows how each organisation helps investigate the ethics suppression in the context of competitive AI development.

| Case | Primary Evidence Sources | Governance Tension Examined | Relevance to Ethics Suppression |
|---|---|---|---|
| **Google** | AI Principles, ethical AI team controversies, governance restructuring, public statements | *Ethical dissent, governance authority, organisational fragmentation* | Illustrates how visible ethics structures may experience weakened operational influence |
| **OpenAI** | Preparedness Framework, governance crisis, safety team departures, public statements | *Safety governance versus deployment acceleration* | Illustrates tensions between responsible AI commitments and competitive scaling pressures |
| **Meta** | Responsible AI documentation, platform governance controversies, Llama deployment strategy | *Rapid deployment, openness, and governance control* | Illustrates prioritisation of strategic expansion over ethical intervention capacity |
| **Microsoft** | Responsible AI Standard, governance frameworks, OpenAI partnership, Copilot deployment | *Governance maturity versus commercial integration pressure* | Illustrates how established governance structures may still experience operational strain under accelerated deployment conditions |

Table 1: Case Selection and Analytical Relevance

The selected cases offer informative examples of organisations that publicly commit to responsible AI governance while facing competitive pressure. Their varied governance structures, organisational models, and deployment strategies enhance the comparative design and highlight common ways ethical oversight can diminish during AI development.

The study focuses on specific cases that serve as insightful examples, highlighting the organisational dynamics of the ethics suppression in competitive AI development, rather than aiming for statistical generalisation across the entire industry.

## Data Collection

The study uses qualitative document analysis of publicly available secondary sources, including corporate AI governance documents, responsible AI principles, safety frameworks, public statements by executives and employees, investigative journalism, governance controversies, academic publications, policy reports, and reputable media coverage on AI governance, safety, and deployment.

These sources highlight organisational responses to ethical concerns, governance disputes, safety oversight, and competitive pressures across the four cases. The analysis aimed to identify common governance patterns by comparing evidence from various documents. Interpretations were formed through cross-case comparisons and source triangulation, and conclusions were drawn only when similar governance dynamics were consistently corroborated by independent sources.

## Analytical Framework

Table 2 details the analytical dimensions used to compare the four cases. These dimensions convert the conceptual framework into an empirical coding structure, enabling the analysis to assess both the existence of responsible AI structures and their effectiveness under competitive pressure.

| Dimension | Purpose in the Analysis |
|---|---|
| Ethics structures | Identifies formal responsible AI mechanisms, principles, review processes, and safety functions. |
| Operational authority | Examines whether ethical oversight could influence development or deployment decisions. |
| Competitive pressure | Assesses evidence of acceleration, market urgency, strategic rivalry, or scaling pressure. |
| Escalation capacity | Examines whether ethical concerns could be raised, reviewed, delayed, or acted upon. |
| Governance outcome | Assesses whether ethical influence was preserved, constrained, bypassed, or weakened. |

Table 2: Cross-Case Analytical Dimensions

The table offers a consistent qualitative framework for comparing ethical oversight across different organisational contexts, rather than imposing a quantitative scoring system.

The analysis takes a process-oriented comparative approach, examining how ethical influences change under various competitive and institutional pressures instead of viewing ethics suppression as a static condition. This method helps identify common organisational mechanisms that undermine or diminish ethical oversight in AI development.

## Data Analysis Procedure

The analysis utilised a structured qualitative document analysis based on the conceptual framework developed in this study. The researcher iteratively reviewed documentary evidence from each organisation, focusing on five analytical dimensions: ethics structures, operational authority, competitive pressure, escalation capacity, and governance outcome, as outlined in Table 2. These dimensions served as coding categories derived from the literature.

The researcher read and coded documentary sources multiple times to identify evidence related to these dimensions. Documents could fit into multiple categories when applicable. The coding process was manual and iterative, comparing evidence within and across the four organisational cases.

To ensure analytical consistency, we triangulated interpretations across various independent sources, including corporate governance documents, responsible AI frameworks, executive statements, investigative journalism, academic publications, and policy reports. We identified cross-case patterns only when multiple sources consistently supported similar governance dynamics, avoiding reliance on isolated events. This iterative comparison between the conceptual framework and the documentary evidence refined our understanding of ethics suppression while maintaining a consistent analytical perspective across all cases.

## Limitations

This study presents some methodological limitations: it relies solely on publicly available secondary sources, including corporate governance documents, public statements, investigative reports, academic literature, and policy materials. While these sources illustrate governance dynamics, they do not provide insights into internal decision-making or confidential discussions. The findings, therefore, indicate observable governance patterns rather than definitive accounts of organisational intent.

The qualitative analysis was conducted by a single researcher. Although guided by a defined framework and a consistent analytical structure, the interpretation is subjective. Different researchers might emphasise different events or governance developments.

Lastly, the study lacks primary data, such as interviews, surveys, or internal records. Future research could enhance the concept of ethics suppression by combining documentary evidence with insights from AI developers, ethics teams, governance professionals, and organisational leaders. This would offer a better understanding of how organisations maintain, negotiate, or weaken ethical authority under competitive pressures.

Despite these limitations, using multiple documentary sources from four major AI organisations provides a solid foundation for theory development. The goal is not statistical generalisation but analytical generalisation through comparative qualitative inquiry, creating a conceptual framework for further empirical research.

## Case Findings: Evidence of Operational Ethics Suppression Across Four Organisations

The analysis reveals a clash between responsible AI governance and rapid development in all four cases. Although the organisations publicly committed to ethical AI, the urgency of strategic needs and competitive pressures limited effective ethical oversight. Each organisation has frameworks for responsible AI and safety, but these structures often fail to uphold ethical standards during high-pressure development phases. Ethics suppression occurred not through outright rejection but through organisational dynamics that diminished the authority of ethical oversight in the development process.

### Google: Governance Visibility and Organisational Fragmentation

Google has been a leader in responsible AI governance since it published its AI Principles in 2018. These principles focus on fairness, accountability, privacy, safety, and the beneficial use of AI. The company has also set up internal structures for responsible innovation and invested significantly in ethical AI research, governance initiatives, and fairness-focused technical development.

Google's commitment to ethical AI was tested by controversies, notably the dismissals of prominent ethics researchers Timnit Gebru and Margaret Mitchell. These incidents sparked public debate about the company's tolerance for ethical dissent and the effectiveness of its internal governance structures (Levy, 2021; Hao, 2020). Critics pointed out that these controversies highlighted a conflict between corporate priorities in AI development and the need for independent ethical oversight, particularly regarding large language models, bias, and social harms.

Concerns about governance arose after reports of restructuring within Google's Responsible Innovation teams and cuts to ethics-focused functions (D’Onfro, 2019). Despite the company's public commitment to responsible AI governance, these changes call into question the stability and effectiveness of ethical oversight amid intensifying competition in AI.

The findings indicate that Google's governance challenges were influenced by its fragmented organisational structure. Ethical oversight was distributed across various units, including engineering, policy, legal, product, and research. While this increased visibility for AI ethics, it also diluted governance authority and complicated the resolution of development disputes.

The case shows that ethical governance exists, but it does not always translate into operational authority in a competitive AI landscape. While institutions can express ethical concerns, these concerns often struggle to influence strategic decisions in rapidly evolving commercial AI settings.

## OpenAI: Competitive Acceleration and Safety Tensions

OpenAI exemplifies the tension between AI safety governance and rapid competitive deployment. While it claims to prioritise long-term AI safety and responsible deployment, it is also a major player in the commercial race for generative AI systems (OpenAI, 2023).

The success of ChatGPT and large language models has elevated OpenAI's role in the global AI market. This growth has led to higher investment expectations, increased pressure for infrastructure expansion, more partnerships, and intensified competition with major tech firms. Consequently, public disputes have arisen over governance, safety oversight, and the pace of deployment.

Several former employees, researchers, and governance observers raised concerns regarding the operational influence of safety oversight within OpenAI's development processes (Mickle et al., 2023; Tan, 2024). Reports on governance restructuring, changes to safety-focused research functions, and the departure of prominent safety researchers sparked broader debate about the operational influence of safety oversight in accelerated-deployment environments (Ropek, 2026). Additionally, the 2023 leadership crisis involving Sam Altman highlighted internal tensions related to governance structures, organisational direction, and the push for faster progress (Mickle & Metz, 2023).

The findings indicate that OpenAI's governance has been increasingly influenced by the pressures of rapid commercialisation of generative AI. While safety governance is still prominent, the urgency to deploy these technologies has limited opportunities for caution, delay, or dissent in the development process.

The case does not outright reject AI ethics or safety governance. Instead, it suggests that increased competition might diminish the effectiveness of ethical oversight mechanisms, even in organisations committed to responsible AI and safety.

## Meta: Strategic Competition and Deployment Prioritisation

Meta has outlined its responsible AI commitments through initiatives focused on fairness, transparency, content governance, and documentation (Meta, 2023). Concurrently, the company has aggressively pursued AI expansion, integrating platforms, deploying generative AI, and developing open models.

The launch of Meta’s Llama models has established the company as a key player in the open-source generative AI space. This strategy has heightened the focus on rapid deployment, ecosystem impact, scalability, and technological positioning in the competitive AI landscape.

Meta's governance has been influenced by ongoing issues related to platform governance, misinformation, algorithmic amplification, and content moderation (Gillespie, 2018). While these challenges existed before the rise of generative AI, they are crucial to understanding the governance tensions associated with large-scale AI deployment.

The findings suggest that Meta’s governance issues stem from prioritising strategic deployment in a competitive platform environment. While responsible AI structures were in place, governance mechanisms often responded to deployment outcomes rather than proactively intervening during development.

Reports on AI-generated misinformation, model-misuse risks, and open-source governance concerns further highlighted tensions between rapid technological expansion and the capacity for ethical oversight (Meta, 2023; Gillespie, 2018). The organisation's focus on scalability and ecosystem expansion has increased the pressure to quickly integrate AI capabilities into its products and services.

The case shows that ethics can be overlooked when institutions prioritise deployment-oriented strategies. While ethical governance structures are still in place, they often take a back seat to goals such as technological growth, market responsiveness, and competitive advantage.

### Microsoft: Institutional Governance and Commercial Integration

Microsoft is a key player in generative AI due to its established governance structures, including a Responsible AI Standard and formal review processes, which were in place before the recent surge in AI deployment (Microsoft, 2024). These measures aim to ensure fairness, reliability, safety, transparency, inclusiveness, and accountability in AI development and use. Additionally, Microsoft's strategic partnership with OpenAI and the swift integration of AI capabilities into products like Copilot highlight its role in the commercial growth of generative AI. This rapid deployment has intensified the need to balance AI feature delivery with governance oversight.

This combination of governance maturity and accelerated commercial deployment makes Microsoft a useful case for examining whether established governance systems remain effective under competitive pressure. Public debates concerning foundation-model risks, AI-generated misinformation, enterprise deployment risks, and governance accountability have highlighted broader concerns regarding the ability of oversight mechanisms to keep pace with rapidly expanding AI capabilities (Bender et al., 2021; Microsoft, 2024).

Unlike the Google and OpenAI cases, evidence of governance conflict at Microsoft is less visible. Nevertheless, the case illustrates an important analytical point: organisations with comparatively developed responsible AI frameworks remain exposed to pressures associated with rapid deployment, commercial integration, and technological competition. This suggests that ethics suppression is not due to a lack of structures, but rather the weakening of their influence under intense deployment incentives.

## Cross-Case Patterns

Recurring patterns were identified in the comparative analysis.

**(a) Ethics Structures Were Institutionally Visible**

All four organisations had clear commitments to responsible AI, established governance frameworks, ethics initiatives, and discussions around public safety. There was no indication of a complete lack of ethical governance structures. This is significant because ethical governance was clearly evident across all organisations, and the ethics suppression did not result from institutional invisibility.

**(b) Operational Authority Was More Limited Than Institutional Visibility**

A key pattern across cases was the difference between ethical visibility and ethical influence. While ethics structures often had advisory or consultative roles and were important for reputation, they lacked the authority to significantly alter strategic decisions in high-pressure situations. This imbalance was consistent across all cases, regardless of variations in organisational structure and governance design.

**(c) Competitive Acceleration Increased Governance Tensions**

In all cases, increased AI competition put greater pressure on organisations to quickly deploy, strategically scale, expand infrastructure, and respond to the market. As a result, ethical review processes often faced shorter timelines and faster decision-making environments. This suggests that competitive pressure may weaken the effectiveness of ethical oversight, even when organisations publicly commit to responsible AI practices.

**(d) Governance Maturity Did Not Eliminate Ethics Suppression Risks**

The case of Microsoft shows that even well-established governance structures can still risk suppressing operational ethics. Despite differences in governance, all organisations dealt with conflicts between ethical oversight and the pressure to deploy quickly. This indicates that the ethics suppression may be a widespread issue in competitive AI environments, not just in poorly governed organisations.

**(e) Ethics Suppression Was Primarily Structural Rather Than Explicit**

The findings did not show a clear rejection of ethics in the organisations studied. Instead, ethics were indirectly suppressed through their structures, incentives, authority structures, and strategic priorities. This is an important analytical distinction. Ethics were weakened not because these organisations publicly abandoned them, but because their environments increasingly prioritised speed, scale, and competitiveness.

### Summary of Findings

The comparative analysis shows that responsible AI governance faces significant operational challenges in competitive AI environments. While organisations publicly uphold strong ethical commitments, tensions arise regarding the authority and timing of ethical oversight in development and deployment.

The findings support the argument that ethical governance structures alone are not sufficient to ensure effective ethical influence in competitive contexts. Instead, the impact of ethical oversight depends on the organisation's authority, its ability to escalate concerns, and the strategic pressures influencing AI deployment decisions.

## Discussion

### From Ethical Presence to Ethical Influence

The findings across the four organisations support the framework shown in Figure 1. Increased competitive pressure was linked to heightened urgency in deployment, which limited escalation mechanisms, reduced the authority of ethical oversight, and led to ethics suppression. Despite variations in governance structures, the overall trend was consistent, indicating that ethics suppression is a recurring governance issue rather than a one-time organisational failure.

The findings highlight a key difference between the presence of AI ethics in institutions and its actual impact on AI development. While existing literature emphasises ethical principles, governance frameworks, transparency, and safety measures (Jobin et al., 2019; Morley et al., 2020), the analysis reveals that having governance mechanisms in place does not necessarily ensure that they influence development and deployment decisions, especially in competitive environments.

In all four cases, ethics were evident in public principles, governance structures, safety commitments, and discussions of responsible AI. However, the effectiveness of ethical oversight was notably limited during times of increased deployment pressure and strategic competition. This demonstrates that the challenges in implementing AI ethics stem not only from poor design but also from organisational factors that undermine its practical authority.

The focus shifts from whether organisations have ethical governance structures to how much influence those structures have in high-pressure development environments. This is crucial because much of the existing AI ethics literature assumes that institutionalisation always improves the effectiveness of governance. Instead, the findings indicate that operational influence is often contingent, unstable, and vulnerable in competitive settings.

## Ethics Suppression as a Structural Organisational Dynamic

The findings indicate that the lack of ethics in AI development stems more from structural issues than from ideological rejection. None of the organisations studied openly dismissed responsible AI principles; instead, all four heavily invested in ethical governance and publicly committed to safety, fairness, accountability, and responsible innovation. But ethical influence was often compromised by organisational factors such as urgency, competition, and strategic rivalry, which prioritised scalability and technological leadership over ethical oversight.

This finding supports existing research on organisational ethics, which shows that ethical concerns can be sidelined by institutional pressures, incentive systems, and a focus on performance, rather than being directly rejected (Tenbrunsel & Messick, 2004). Similar trends are observed in studies of organisational silence and ethical fading, in which moral considerations diminish in environments focused on strategic or performance goals (Bird & Waters, 1989).

In AI development environments, the competition around generative AI is intensifying. The urgent pace of development often clashes with governance processes that typically involve careful discussion and review, prioritising quick deployment and market responsiveness instead.

## Extending the Ethics Washing Literature

The findings contribute to the ongoing discussions surrounding ethics washing in the governance of artificial intelligence. Ethics washing refers to circumstances in which organisations adopt ethical language, principles, or governance frameworks primarily to convey a sense of responsibility without implementing meaningful changes in organisational behaviour (Bietti, 2020). Existing research has largely focused on the dissonance between external ethical representations and actual organisational practices.

The present study identifies a related yet distinct phenomenon. The organisations examined did not merely demonstrate symbolic ethical commitments. All four case studies demonstrated robust responsible AI frameworks, including governance structures, safety initiatives, review mechanisms, and public commitments to ethical AI practices. Therefore, the central issue is not the lack of governance structures or the superficial adoption of ethical principles, but rather the depth and effectiveness of their implementation.

The findings indicate that ethical governance may remain embedded within institutions while facing a reduction in its influence over decisions related to development and deployment. It is important to differentiate ethics suppression from ethics washing; the former focuses on the operational authority of ethical oversight rather than the authenticity of ethical commitments. The key analytical question is not whether organisations genuinely endorse responsible AI, but whether ethical governance continues to exert sufficient influence amid escalating strategic and competitive pressures.

This distinction contributes to the existing literature on ethics washing by introducing a governance perspective that emphasises authority over mere symbolism. While ethical commitments may be sincere, governance structures may be formally established, and responsible AI initiatives may be actively pursued, ethical oversight can still experience diminishing operational influence if organisational incentives increasingly prioritise rapid deployment, scalability, and competitive responsiveness. Consequently, ethics suppression serves as a complementary explanation for the limitations of governance within today's AI development environments, rather than a replacement for ethics washing.

## Competitive Pressure and the Compression of Ethical Deliberation

A key finding was the link between competitive pressure and shorter governance timelines. Increased AI competition shortened governance periods, heightened deployment urgency, and limited opportunities for ethical discussions.

This trend poses significant challenges for responsible AI governance. Ethical review processes typically require extensive time for interdisciplinary consultations, risk assessments, and institutional coordination. With faster development cycles, these processes may shift from being meaningful to becoming merely procedural or reactive.

AI competition is likely to change both technological development and the conditions for ethical governance. As organisations strive for market leadership and investment, governance structures may struggle to keep up with rapid deployment. This could undermine the effectiveness of ethical oversight, making it more influenced by commercial interests.

The issue is important because competition in generative AI is similar to strategic technological races in other fast-paced innovation sectors. Historical research has shown that intense rivalry can change how organisations manage risk, govern themselves, and make decisions (Christensen, 1997). This suggests that AI governance may face similar structural challenges.

## Implications for Responsible AI Governance

Current governance discussions often focus on ethical principles, technical safeguards, transparency, and regulatory compliance. While these aspects are important, the effectiveness of governance also relies significantly on organisational authority structures and their ability to escalate issues.

Responsible AI governance should be more integrated into development decision-making rather than merely advisory. Ethical oversight without operational authority may lose its influence during times of rapid change. Additionally, simply having ethics principles, safety policies, or governance committees is not enough to measure governance effectiveness. Organisations can have strong responsible AI frameworks yet exhibit weak operational ethics under competitive pressure.

Regulators, policymakers, and governance scholars should pay attention. Governance assessment frameworks need to clearly focus on:

• escalation authority

• deployment veto capacity

• independence of safety review structures

• organisational reporting lines

• ability of ethics mechanisms to affect strategic decisions under pressure

Without operational authority, ethical governance may appear visible in process but be limited in effectiveness.

### Theoretical Implications

This paper enhances the responsible AI governance literature by introducing the concept of ethics suppression, which is influenced by competitive and institutional pressures. This shifts the focus from mere ethical representation to the operational impact of ethics in AI development.

The findings also provide insights into how intense technological competition can disrupt the conditions needed for ethical decision-making. AI development environments are particularly vulnerable to governance compression due to the simultaneous effects of rapid technological advancements, commercial interests, and strategic competition.

Overall, ethical governance should not be seen as a fixed achievement. Instead, it is dynamic and influenced by changing organisational conditions. The effectiveness of governance can vary with competitive intensity, strategic urgency, and organisational structure, rather than merely the presence of ethical principles or frameworks.

## Theoretical Contributions

### Extending Responsible AI Governance Literature

The paper shifts focus from creating ethical frameworks to examining the authority of ethical oversight in AI development organisations. It emphasises that responsible AI governance goes beyond having principles or committees; it examines whether governance structures can meaningfully influence deployment decisions, especially in urgent or competitive situations. This perspective adds to existing governance research by highlighting when ethical influence is maintained or weakened in practice.

The study challenges the assumption that ethical governance always has a strong influence. It shows that while ethical structures may still be visible, their effectiveness can weaken in competitive environments. The findings enhance our understanding of responsible AI governance by emphasising that governance should be viewed not just as the existence of ethical principles or review boards, but as the real ability of oversight mechanisms to influence decisions in challenging circumstances.

This contribution highlights that current governance discussions often focus too much on design while neglecting the organisational factors that influence governance authority. The study finds that the effectiveness of ethical governance relies not only on governance structure but also on escalation capacity, institutional independence, authority distribution, and the organisation's ability to manage delays, dissent, and caution during rapid innovation.

The paper emphasises a shift in AI governance scholarship from principle-based ethics to the analysis of operational governance. It stresses the importance of understanding how governance structures perform under real organisational pressures, rather than assuming that ethical institutionalisation guarantees effective oversight.

## Extending Organisational Ethics Literature

This paper advances the literature on organisational ethics by applying concepts such as ethical fading, organisational silence, and moral disengagement to AI development environments (Bird & Waters, 1989; Tenbrunsel & Messick, 2004). Current research shows that ethical concerns often become less visible in organisations influenced by performance pressure, strategic incentives, and hierarchy. However, these issues have been explored only minimally in the context of large-scale AI development and technological competition.

AI development environments create unique conditions that can intensify ethics suppression. Unlike traditional organisations, AI companies face rapid technological change, intense competition, significant investment pressures, and a rush for first-mover advantage. These factors can shorten governance timelines and limit the space for ethical discussions.

This study enhances our understanding of organisational ethics by demonstrating how fast-paced technological competition alters the institutional context for ethical considerations. Ethics suppression in AI development isn't always due to overt opposition to ethics; rather, ethical concerns may become less relevant in environments that prioritise speed, scalability, and responsiveness.

The contribution to organisational ethics shows that ethical influence can diminish even when formal commitments are upheld. In competitive AI environments, ethical concerns are not outright rejected but are increasingly sidelined due to tight deadlines, fragmented authority, and a focus on deployment incentives. This shift highlights the need to examine the organisational conditions that impact the relevance of ethical considerations in decision-making processes.

## The Operational Suppression Lens for Analysing Ethical Influence

The paper introduces an operational-suppression lens for analysing ethical governance in AI development. While existing literature on ethics washing emphasises the disconnect between ethical representation and genuine organisational commitment (Bietti, 2020), it fails to address cases where ethical governance structures exist but become ineffective during development processes.

The operational suppression lens examines how internal organisational dynamics can weaken or marginalise ethical concerns under competitive pressures. It shifts the focus from whether organisations publicly endorse ethics to whether ethical oversight has real authority in decision-making. This lens does not suggest that organisations intentionally abandon ethics or deceive governance. Instead, it views the ethics suppression as a structural process arising from factors such as competition, incentives, fragmented governance, the distribution of authority, and time constraints.

This perspective offers a process-oriented view of the decline of governance in AI organisations. Ethical influence is seen as a dynamic condition, not a fixed trait, shaped by evolving strategic circumstances. As a result, governance structures may seem stable while simultaneously losing operational impact due to increased competitive pressure.

The operational suppression lens adds value by providing a framework for analysing governance effectiveness beyond visible policies and structures. Instead of only looking at responsible AI through policies, ethics principles, or governance committees, this framework promotes a broader examination. It shifts the focus of governance analysis from simply having ethics programs to examining how ethical influence is applied or limited. It therefore encourages assessment of:

- operational authority
- escalation effectiveness
- deployment influence
- governance independence
- institutional capacity to resist acceleration pressures

## Toward a Dynamic Theory of Ethical Influence in AI Development

The paper argues that ethical governance in AI development is dynamic and vulnerable, rather than simply secured by formal frameworks. It shows that the ethical influence varies with organisational conditions, strategic pressures, authority structures, and competitive environments. Effective AI governance relies on the ability of ethical oversight mechanisms to maintain authority amid rapid innovation and competition. By introducing the concept of ethics suppression, the paper provides a new perspective on the relationship between AI governance, technological competition, and ethical authority in today's AI development landscape.

## Governance Implications and Practical Considerations

### From Ethical Visibility to Operational Authority

The effectiveness of responsible AI governance relies more on its ability to influence development outcomes than on the visibility of ethics structures. In the cases examined, governance mechanisms were present but varied significantly in their operational impact under competitive pressure. Therefore, governance should be assessed not just by the existence of oversight structures but by their ability to influence deployment decisions when incentives for acceleration are present.

This distinction is crucial for designing AI governance. Ethics structures that serve mainly as advisory or reputational tools may struggle to influence deployment decisions under strong commercial and strategic pressures. Effective AI governance needs institutional mechanisms that can influence development timelines, deployment choices, and escalation processes.

Organisations should assess responsible AI governance not just by the presence of governance structures but also by the effectiveness and quality of those structures.

• escalation authority

• institutional independence

• deployment influence

• operational integration into strategic decision-making processes

Without authority, ethical governance may seem visible in process but is limited in its effectiveness.

### Strengthening Escalation and Review Capacity

The analysis highlights the need for robust escalation mechanisms in AI governance systems. In the cases examined, ethical concerns often fell short when governance lacked the power to delay or challenge significant deployment decisions. This indicates that effective responsible AI governance requires stronger independent mechanisms, especially in fast-paced development environments. Consultation-focused governance may not be sufficient under such pressure.

Governance scholars argue that responsible AI requires institutional accountability structures rather than merely voluntary ethical signalling (Metcalf et al., 2019; Raji et al., 2020). These findings support this view, indicating that effective governance relies on the ability of ethical oversight to intervene in development processes when concerns arise.

Practical governance mechanisms can include:

• independent review pathways for high-risk systems

• formal escalation procedures for unresolved ethical concerns

• governance structures capable of delaying deployment when necessary

• clearer separation between commercial product incentives and safety oversight functions

The goal is to maintain the pace of innovation while ensuring that ethical oversight can influence important development decisions, even under competitive pressure.

## Embedding Governance Within Development Processes

The findings indicate that responsible AI governance can be ineffective if it is seen mainly as a compliance or post-development process. In many instances, ethical governance appears reactive rather than fully integrated into core development cycles. This observation supports existing research suggesting that ethical mechanisms are more effective when incorporated directly into organisational workflows rather than applied externally or after the fact (Morley et al., 2020). Responsible AI governance, therefore, may require deeper integration into:

• model development

• testing environments

• deployment approvals

• infrastructure scaling decisions

• product release processes.

In AI environments where rapid iteration and continuous deployment are standard, embedding governance within development cycles is crucial. If governance processes operate separately from development workflows, they may lose their effectiveness. Therefore, it's important to adopt governance-by-design approaches, integrating ethical oversight into the organisation's routine operations rather than relying solely on external reviews or reputational assurances.

## Temporal Governance and the Protection of Ethical Deliberation

A key finding from the comparative analysis is that rapid deployment cycles can weaken ethical oversight by limiting time for interdisciplinary review, debate, and risk assessment. This poses significant challenges for AI governance, as responsible AI practices rely on thorough processes that require time for technical, ethical, and legal evaluations, as well as cross-functional collaboration. With compressed timelines, governance tends to follow procedures rather than engage in meaningful deliberation. This highlights the need for "temporal governance," which focuses on preserving ethical discussions despite accelerated development.

Temporal governance may include:

• mandatory review intervals for high-impact systems

• staged deployment checkpoints

• minimum testing and evaluation periods

• governance pause mechanisms

• protected escalation windows for unresolved ethical concerns

As AI development shifts toward continuous deployment and rapid commercial integration, effective governance becomes essential. Without protected timelines for governance, ethical reviews may diminish in impact, even if formal structures are still in place.

## Governance Resilience Under Competitive Pressure

The main takeaway from this study is that effective AI governance is truly tested not during stable conditions, but in times of intense competition. While governance systems may seem effective during periods of slower development, they often weaken as urgency and strategic pressures increase.

To evaluate responsible AI governance, it is essential to consider its resilience in high-pressure situations rather than just checking for ethical principles and documentation. This approach focuses on the practical capabilities of governance systems rather than their static institutional appearances.

The findings suggest that governance resilience is highly dependent on:

• independence of oversight structures

• authority of safety escalation pathways

• leadership tolerance for delay and dissent

• organisational willingness to preserve ethical influence during competitive acceleration

As competition in generative AI grows, governance systems will be judged on their ability to uphold ethical standards amidst rapid development. The main challenge is not just adopting responsible AI principles but ensuring that institutions can maintain authority when faced with pressure to deploy quickly. The effectiveness of governance depends on the ability to influence important decisions, especially when delaying could incur significant costs.

### Organisational Costs of Ethics Suppression

The findings indicate that ethics suppression in AI development creates a strategic paradox. Organisations may gain from quicker deployment, fewer governance hurdles, faster decision-making, and improved market responsiveness. These advantages can enhance market positioning and enable firms to capitalise on first-mover benefits in rapidly evolving AI markets.

However, these short-term gains may mask high long-term costs. As ethical oversight weakens, organisations may face increased risks of governance failures, regulatory actions, litigation, reputational harm, and diminished stakeholder trust. A lack of ethical authority can impair the organisation's ability to detect and manage emerging risks, allowing issues related to safety, fairness, transparency, or accountability to persist until deployment. Such issues can lead to higher compliance costs, necessitate expensive product changes or withdrawals, and threaten organisational legitimacy in increasingly regulated AI environments (Raji et al., 2020; Morley et al., 2020; NIST, 2023).

The findings also highlight that ethics suppression can lead to a buildup of unresolved ethical risks that extend beyond individual development cycles, exposing organisations to ongoing operational, legal, financial, and societal repercussions. Thus, the supposed efficiency gains from weakened ethical oversight should be viewed as potential trade-offs rather than sustainable competitive advantages. This reinforces the argument that effective responsible AI governance relies not only on ethical frameworks but also on the operational authority to enforce them in competitive environments.

## Conclusion

This study set out to explain how ethics suppression emerges as an organisational governance phenomenon in competitive AI development environments and to develop a conceptual framework for analysing how competitive pressure influences the practical authority of ethical oversight. Through a comparative qualitative analysis of Google, OpenAI, Meta, and Microsoft, the study demonstrates that organisations may continue to maintain visible responsible AI frameworks, ethics principles, and governance structures while simultaneously experiencing a decline in the practical influence of ethical oversight over development and deployment decisions.

The principal theoretical insight of the study is that the effectiveness of responsible AI governance should not be assessed solely by the existence of formal ethical frameworks or governance mechanisms. Rather, governance effectiveness depends on whether ethical oversight retains sufficient operational authority to influence organisational decisions under conditions of competitive pressure. This shifts attention from the institutional design of responsible AI governance towards the organisational conditions that determine whether ethical oversight remains practically influential during AI development.

Building on this insight, the study introduces ethics suppression as a distinct organisational governance phenomenon, distinct from ethics washing. Whereas ethics washing primarily concerns the external projection of ethical responsibility, ethics suppression explains the internal organisational processes through which ethical oversight may gradually lose influence despite remaining institutionally visible. The comparative findings indicate that competitive pressure, deployment urgency, weakened escalation capacity, fragmented governance, and reduced operational authority can collectively diminish the practical effectiveness of responsible AI governance.

The study further proposes the operational suppression lens as a complementary framework for evaluating responsible AI governance. Rather than asking whether organisations possess ethics principles, governance committees, or responsible AI policies, the lens focuses on whether these mechanisms retain meaningful authority over development and deployment decisions when commercial and strategic pressures intensify. This perspective provides researchers with a new analytical approach to examining governance effectiveness and offers organisations and policymakers a practical framework for assessing whether ethical oversight remains operationally effective amid accelerated AI innovation.

The findings demonstrate that institutional visibility and operational authority should not be treated as equivalent indicators of effective AI governance. Organisations may publicly demonstrate strong commitments to responsible AI through ethics frameworks, governance structures, and transparency initiatives while simultaneously experiencing a weakening of the practical authority needed for ethical oversight to shape organisational decisions. Recognising this distinction provides a more realistic basis for evaluating responsible AI governance and helps explain why governance failures may emerge even in organisations that appear institutionally committed to responsible AI.

As AI competition continues to accelerate, the challenge for organisations extends beyond establishing ethical principles or governance structures. The more fundamental challenge is ensuring that ethical oversight retains sufficient authority to influence decision-making when commercial, technological, and competitive pressures are greatest. Ultimately, the long-term credibility of responsible AI governance will depend not only on its institutional visibility but also on its continued operational authority in practice.